\documentclass[%
reprint,
superscriptaddress,
nofootinbib,
amsmath,
amssymb,
aps,
floatfix,
showkeys,
]{revtex4-2}

\usepackage[colorlinks,citecolor=blue,linkcolor=blue,anchorcolor=blue,filecolor=blue,
urlcolor=blue]{hyperref}

\usepackage{graphicx,color,xcolor}
\usepackage{amsfonts,amsmath}
\usepackage{amssymb,ulem}
\usepackage{dcolumn}
\usepackage{bm,lipsum}
\usepackage[utf8]{inputenc}
\usepackage{subeqnarray}
\usepackage{array}
\usepackage{epsfig,hyperref}
\usepackage{morefloats}
\usepackage{relsize}
\usepackage{url}
\usepackage{orcidlink}
\usepackage{lineno}

\usepackage{comment}

\begin{document}

\preprint{ }

\title{In-medium effects and thermodynamic consistency of non-constant bag models}

\author{Luiz L. Lopes~\orcidlink{0000-0003-0982-9774}}

\affiliation{ Centro Federal de Educac\~{a}o Tecnol\'{o}gica de Minas Gerais Campus VIII; CEP 37.022-560, Varginha - MG - Brazil\\}

\begin{abstract}
Thermodynamic consistency is not always explicitly verified in recent applications of non-constant bag models. Here, I derive consistent quark-matter equations of state for density-dependent and chemical-potential-dependent bag functions. This leads to the onset of rearrangement terms that modify chemical potentials and pressure in the former case, and number and energy densities in the latter. After explicitly verifying that both formulations satisfy the first law of thermodynamics and the Euler relation, I investigate their stellar predictions using a Woods--Saxon-type parametrization. I find that omitting the rearrangement terms can overestimate maximum stellar masses. A consistent color-flavor-locked model reaches $M_{\max}=2.33\,M_\odot$ and predicts $\Lambda_{1.4}=578$, satisfying the adopted constraints from PSR J0740+6620 and GW170817 while recovering the conformal limit, $v_s^2\rightarrow1/3$, at asymptotically high densities.
\end{abstract}

\maketitle

\section{Introduction}

The behavior of strongly interacting matter at extreme densities remains one of the major open questions in modern physics. In particular, the composition of matter in the inner regions of massive neutron stars is still uncertain, with recent studies suggesting that their cores may contain deconfined quark matter~\cite{nature_2020,Annala2023}. A more extreme possibility arises from the Bodmer--Witten conjecture, according to which strange quark matter may represent the absolute ground state of strongly interacting matter~\cite{Bodmer,Witten}. If this conjecture is true, neutron stars may undergo a conversion into strange stars, composed entirely of self-bound quark matter~\cite{Olinto}. Understanding the equation of state of quark matter is therefore essential for investigating both possibilities.

One possible approach to obtaining sufficiently stiff equations of state for quark matter, capable of satisfying modern astrophysical constraints such as those imposed by the massive pulsar PSR J0740+6620~\cite{Salmi_2024}, is to extend the MIT bag model by incorporating repulsive vector interactions among quarks. Such interactions increase the pressure at high densities, allowing the resulting equations of state to support massive strange stars~\cite{lopesps1,lopesps2}. This approach has been successfully employed to describe both self-bound strange stars and hybrid stars containing deconfined quark cores~\cite{Lopes2022ApJ,lopesEPJC2025}.

Despite these successes, the conventional vector MIT bag model presents important limitations when extrapolated to asymptotically high densities. One concern is the treatment of the QCD vacuum, which is represented by a constant bag pressure independent of the baryon density. In QCD, the strong coupling $\alpha_s$ runs with the energy scale and decreases as the characteristic quark momenta increase, reflecting asymptotic freedom~\cite{Buballa2005}. Although the relative contribution of a constant bag becomes negligible in this limit, a fixed bag parameter does not describe the possible evolution of nonperturbative vacuum effects in the dense medium. Another limitation arises from the repulsive vector interaction, which generally produces a speed of sound that increases with density, in contrast to the perturbative QCD prediction that $v_s^2$ approaches the conformal limit of $1/3$ at asymptotically high densities~\cite{Kurkela2010}.

One possible way to overcome some of these limitations is to introduce a density-dependent bag parameter that decreases as the baryon density increases. This approach allows the effective bag pressure to gradually ``melt'' at high densities, suppressing the vacuum contribution associated with confinement and providing a phenomenological description of medium effects in deconfined quark matter. Furthermore, a decreasing bag parameter can lead to a stiffer equation of state and consequently support more massive compact stars than those predicted by the corresponding constant-bag model.

Such models have been explored in several previous studies. However, their thermodynamic treatment has not always been carried out consistently. In particular, the conventional MIT bag model expressions have sometimes been employed without accounting for the additional derivative contributions arising from the density dependence of the bag parameter~\cite{Sen2021,Pal2023,Podder2024}. Such an omission leads to thermodynamic inconsistencies, violating the differential thermodynamic relations even when the Euler relation remains algebraically satisfied, and may artificially increase the predicted pressure, thereby overestimating the maximum masses of compact stars.

In other, more elaborate applications, including perturbative QCD and magnetized quark matter, an explicit verification of both the first law and the Euler relation is not presented~\cite{Sedaghat2025,Bordbar2012}. In particular, Ref.~\cite{Sedaghat2025} combines a thermodynamic potential expressed in terms of chemical potentials with a bag function prescribed in terms of the baryon density. Such a formulation requires a consistent treatment of the implicit density dependence when evaluating thermodynamic derivatives.

The aim of the present work is to investigate bag models in which the in-medium effects are explicitly taken into account, ensuring full thermodynamic consistency. I begin with the conventional MIT bag model extended to include a density-dependent bag parameter, as previously explored in the literature~\cite{Sen2021,Pal2023,Podder2024}. In this formulation, additional rearrangement terms naturally emerge from the density dependence of the bag function and must be properly incorporated into the thermodynamic quantities to preserve both the first law of thermodynamics and the Euler relation.

I then generalize this treatment to color-superconducting quark matter, whose thermodynamic description is naturally formulated in terms of chemical potentials~\cite{Alford2005,Lugones2002,Pereira2018}. In particular, I construct a color-flavor-locked (CFL) model with a chemical-potential-dependent bag parameter, deriving a fully thermodynamically consistent equation of state that accounts for all additional contributions arising from this dependence. For both models, I also calculate the equations of state neglecting the corresponding rearrangement terms, allowing me to quantitatively assess their impact on the thermodynamic properties and the resulting predictions for compact stars.

\section{Density-dependent bag models}
\label{sec:formalism}

The bulk Lagrangian of the MIT bag model reads~\cite{Chodos1974,Johnson1978}:

\begin{equation}
\mathcal{L}=\left[\sum_q\bar{\psi}_q
\left(i\gamma^\mu\partial_\mu-m_q\right)\psi_q-B(n)\right]\Theta(\bar{\psi}_q\psi_q),
\label{eq:lagrangian}
\end{equation}
where the sum runs over the three lightest quark flavors ($q=u,d,s$), $\psi_q$ is the quark field, $B(n)$ is an arbitrary differentiable function of the baryon number density $n$, and $\Theta(\bar{\psi}_q\psi_q)$ is the Heaviside step function defining the interior of the bag. Only the bulk contribution is considered here; surface contributions are not relevant to homogeneous infinite matter, for which $\Theta(\bar{\psi}_q\psi_q)=1$. In the rest frame of the matter, the baryon density is given by:

\begin{equation}
 n=\frac{1}{3}\sum_q n_q
 =\frac{1}{3}\sum_q\bar{\psi}_q\gamma^0\psi_q,
\label{eq:density_field}
\end{equation}
where $n_q$ is the number density of quarks of flavor $q$. The factor $1/3$ accounts for the baryon number carried by each quark. In the mean-field treatment, the density is replaced by its expectation value.

Applying the Euler--Lagrange equation to the adjoint quark field, one obtains:

\begin{equation}
\frac{\partial\mathcal{L}}{\partial\bar{\psi}_q}
-\partial_\mu\left[\frac{\partial\mathcal{L}}
{\partial(\partial_\mu\bar{\psi}_q)}\right]=0,
\label{eq:EL}
\end{equation}
and  the equation of motion in the interior of the bag becomes:

\begin{equation}
\left(i\gamma^\mu\partial_\mu-m_q\right)\psi_q
-\frac{dB}{dn}\frac{\partial n}{\partial\bar{\psi}_q}=0.
\label{eq:EL_chain}
\end{equation}

From Eq.~\eqref{eq:density_field}:

\begin{equation}
\frac{\partial n}{\partial\bar{\psi}_q}
=\frac{1}{3}\gamma^0\psi_q.
\end{equation}

Thus, the modified Dirac equation reads:
\begin{equation}
\left[i\gamma^\mu\partial_\mu-m_q
-\frac{1}{3}\frac{dB}{dn}\gamma^0\right]\psi_q=0.
\label{eq:Dirac}
\end{equation}

For homogeneous matter in the mean-field approximation, plane-wave solutions can be written as $\psi_q(x)=u_q(\bm{k})\exp(-ik_\mu x^\mu)$, where $k^\mu=(E_q,\bm{k})$. Multiplying Eq.~\eqref{eq:Dirac} by $\gamma^0$, one obtains:

\begin{equation}
\left(\bm{\alpha}\cdot\bm{k}+\beta m_q\right)u_q
=\left(E_q-\frac{1}{3}\frac{dB}{dn}\right)u_q,
\end{equation}
where $\bm{\alpha}=\gamma^0\bm{\gamma}$ and $\beta=\gamma^0$~\cite{GreinerRQM}. The energy eigenvalue is therefore:

\begin{equation}
 E_q(k)=\sqrt{k^2+m_q^2}+\frac{1}{3}\frac{dB}{dn}.
\label{eq:eigenvalue}
\end{equation}

The last term in the expression above can be recognized as a rearrangement term, as originally introduced in Ref.~\cite{TYPEL1999}. We can write:

\begin{equation}
 \Sigma_R\equiv\frac{1}{3}\frac{dB}{dn},
\label{eq:rearrangement}
\end{equation}
so that $E_q(k)=\sqrt{k^2+m_q^2}+\Sigma_R$. Notice that the rearrangement term emerges naturally from the Euler--Lagrange equation and is not introduced \textit{ad hoc}.

At zero temperature, the Fermi--Dirac distribution reduces to a step function.
In the no-sea approximation, the quark number density is
\begin{equation}
 n_q=\frac{\gamma_q}{2\pi^2}\int_0^{k_{F,q}}k^2\,dk
 =\frac{\gamma_q k_{F,q}^3}{6\pi^2},
\label{eq:fermi_density}
\end{equation}
where $\gamma_q=6$ accounts for spin and color degeneracy. From Eq.~\eqref{eq:fermi_density}:

\begin{equation}
 \frac{dn_q}{dk_{F,q}}=\frac{\gamma_q k_{F,q}^2}{2\pi^2},
 \qquad
 \frac{dk_{F,q}}{dn_q}=\left(\frac{dn_q}{dk_{F,q}}\right)^{-1}.
\label{eq:density_derivatives}
\end{equation}
The inverse derivative is understood for $k_{F,q}>0$. The baryon density and its partial derivative are then:

\begin{equation}
 n=\sum_q\frac{\gamma_q k_{F,q}^3}{18\pi^2},\qquad
 \frac{\partial n}{\partial k_{F,q}}=\frac{\gamma_q k_{F,q}^2}{6\pi^2},
\label{eq:baryon_momentum}
\end{equation}
where the other Fermi momenta are held fixed in the partial derivative. The quark chemical potential is the single-particle energy evaluated at the Fermi surface:
\begin{equation}
 \mu_q=E_q(k_{F,q})=\sqrt{k_{F,q}^2+m_q^2}+\Sigma_R.
\label{eq:muq}
\end{equation}

The energy density can now be obtained directly from the first law of thermodynamics at zero temperature:

\begin{equation}
 d\varepsilon=\sum_q\mu_q\,dn_q.
\label{eq:first_law}
\end{equation}
Using Eqs.~\eqref{eq:muq} and \eqref{eq:density_derivatives}, the first law can be expressed entirely in terms of the Fermi momenta:
\begin{align}
 d\varepsilon={}&\sum_q\frac{\gamma_q k_{F,q}^2}{2\pi^2}
 \left(\sqrt{k_{F,q}^2+m_q^2}+\Sigma_R\right)dk_{F,q}\notag\\
 ={}&\sum_q\frac{\gamma_q k_{F,q}^2}{2\pi^2}
 \sqrt{k_{F,q}^2+m_q^2}\,dk_{F,q}+dB,
\label{eq:de}
\end{align}
where, by Eqs.~\eqref{eq:rearrangement} and \eqref{eq:baryon_momentum}:

\begin{equation}
 dB=\sum_q\frac{\partial B}{\partial k_{F,q}}\,dk_{F,q},\quad
 \frac{\partial B}{\partial k_{F,q}}=
 \frac{\gamma_q k_{F,q}^2}{2\pi^2}\Sigma_R.
\label{eq:bag_momentum_derivative}
\end{equation}
Integrating Eq.~\eqref{eq:de} from the empty bag to the state with Fermi momenta $k_{F,q}$ gives:

\begin{align}
 \varepsilon(n)-\varepsilon(0)={}&
 \sum_q\frac{\gamma_q}{2\pi^2}\int_0^{k_{F,q}}k^2\sqrt{k^2+m_q^2}\,dk\notag\\
 &+B(n)-B(0).
\label{eq:energy_definite}
\end{align}

In the absence of quarks, the Lagrangian density inside the bag reduces to $-B(0)$. With the vacuum energy outside the bag taken as the reference, the energy density inside an empty bag is therefore $\varepsilon(0)=B(0)$. Consequently, the energy density becomes:
\begin{equation}
 \varepsilon=\sum_q\frac{\gamma_q}{2\pi^2}
 \int_0^{k_{F,q}}k^2\sqrt{k^2+m_q^2}\,dk+B(n).
\label{eq:energy}
\end{equation}

Although the rearrangement term explicitly modifies the single-particle energies, there is no additional rearrangement contribution to the total energy density beyond $B(n)$ itself. 
The pressure is obtained directly from the derivative of the energy per baryon. For a compression at fixed composition, I define $Y_q=n_q/n$, with $\sum_q Y_q=3$, and write
\begin{equation}
 p=n^2\left.\frac{\partial(\varepsilon/n)}{\partial n}\right|_{\{Y_q\}}
 =n\left.\frac{\partial\varepsilon}{\partial n}\right|_{\{Y_q\}}-\varepsilon.
\label{eq:pressure_def}
\end{equation}

Since $n_q=Y_q n$, the derivative of the Fermi momentum follows directly from Eq.~\eqref{eq:density_derivatives}:
\begin{align}
 \left.\frac{\partial k_{F,q}}{\partial n}\right|_{\{Y_q\}}
 &=Y_q\left(\frac{dn_q}{dk_{F,q}}\right)^{-1}\notag\\
 &=\frac{2\pi^2Y_q}{\gamma_q k_{F,q}^2}
 =\frac{k_{F,q}}{3n}.
\label{eq:k_density_derivative}
\end{align}

An absent flavor contributes zero to the compression. The chain rule therefore gives the following useful identity for any differentiable function $F$ of the Fermi momenta:
\begin{equation}
 n\left.\frac{\partial F}{\partial n}\right|_{\{Y_q\}}
 =\frac{1}{3}\sum_q k_{F,q}\frac{\partial F}{\partial k_{F,q}}.
\label{eq:momentum_operator}
\end{equation}
In particular, the pressure can be written entirely in terms of derivatives with respect to the Fermi momenta:
\begin{equation}
 p=\frac{1}{3}\sum_q k_{F,q}
 \frac{\partial\varepsilon}{\partial k_{F,q}}-\varepsilon.
\label{eq:pressure_momentum}
\end{equation}
Differentiating Eq.~\eqref{eq:energy} with respect to each Fermi momentum and using Eq.~\eqref{eq:bag_momentum_derivative}, I obtain:

\begin{equation}
 \frac{\partial\varepsilon}{\partial k_{F,q}}=
 \frac{\gamma_q k_{F,q}^2}{2\pi^2}
 \left(\sqrt{k_{F,q}^2+m_q^2}+\Sigma_R\right).
\label{eq:denergy}
\end{equation}

Equivalently, Eq.~\eqref{eq:momentum_operator} yields:
\begin{equation}
 \left.\frac{\partial\varepsilon}{\partial n}\right|_{\{Y_q\}}
 =\sum_qY_q\left(\sqrt{k_{F,q}^2+m_q^2}+\Sigma_R\right).
\label{eq:dedn}
\end{equation}
Substituting Eq.~\eqref{eq:denergy} into Eq.~\eqref{eq:pressure_momentum}, the pressure becomes:

\begin{align}
 p={}&\sum_q\frac{\gamma_q}{2\pi^2}\bigg[
 \frac{k_{F,q}^3}{3}\sqrt{k_{F,q}^2+m_q^2}\notag\\
 &\qquad-\int_0^{k_{F,q}}k^2\sqrt{k^2+m_q^2}\,dk\bigg]\notag\\
 &-B(n)+3n\Sigma_R.
\label{eq:pressure_intermediate}
\end{align}
Integrating by parts,
the pressure finally reads:

\begin{equation}
 \begin{aligned}
 p={}&\sum_q\frac{\gamma_q}{6\pi^2}\int_0^{k_{F,q}}
 \frac{k^4}{\sqrt{k^2+m_q^2}}\,dk\\
 &-B(n)+3n\Sigma_R.
 \end{aligned}
\label{eq:pressure}
\end{equation}
Unlike the energy density, the pressure contains an explicit rearrangement contribution, $3n\Sigma_R=n\,dB/dn$.

To verify the thermodynamic consistency of the model, I now evaluate the sum of the explicit integral expressions for the energy density and pressure without invoking the Euler relation. From Eqs.~\eqref{eq:energy} and \eqref{eq:pressure}:
\begin{align}
 \varepsilon+p={}&\sum_q\frac{\gamma_q}{6\pi^2}
 \int_0^{k_{F,q}}\bigg[3k^2\sqrt{k^2+m_q^2}\notag\\
 &\qquad+\frac{k^4}{\sqrt{k^2+m_q^2}}\bigg]dk+3n\Sigma_R.
\label{eq:enthalpy_integral}
\end{align}

The integrand is precisely the derivative of $k^3\sqrt{k^2+m_q^2}$ with respect to $k$. Hence,
\begin{align}
 \varepsilon+p&=\sum_q\frac{\gamma_q k_{F,q}^3}{6\pi^2}
 \sqrt{k_{F,q}^2+m_q^2}+3n\Sigma_R\notag\\
 &=\sum_q n_q\left(\sqrt{k_{F,q}^2+m_q^2}+\Sigma_R\right)\notag\\
 &=\sum_q n_q\mu_q.
\label{eq:Euler_q}
\end{align}

Therefore, the first law and the Euler relation are simultaneously satisfied for an arbitrary differentiable bag function $B(n)$. If the rearrangement terms are neglected, the Euler relation is still satisfied algebraically, but this prescription yields $d\varepsilon/dn-\mu_B=dB/dn$ and $n^2d(\varepsilon/n)/dn-p=n\,dB/dn$. Thus, both the first law and the thermodynamic relation between pressure and energy density are violated whenever $dB/dn\neq0$. In the density-independent limit, $\Sigma_R=0$, and all conventional MIT bag model expressions are recovered.

\subsection{A simple case}
\label{sec:simple}

As a simple illustration of the formalism, I adopt a Woods--Saxon-type functional form~\cite{Woods1954} for the density dependence of the bag parameter:

\begin{equation}
 B(n)=\frac{B_0}{1+\exp\left(\dfrac{n-n'}{\Delta n}\right)}.
\label{eq:woods_saxon}
\end{equation}
Here, $B_0>0$ sets the amplitude, $n'$ is the density at which $B(n')=B_0/2$, and $\Delta n>0$ controls the diffuseness of the transition. From Eq.~\eqref{eq:rearrangement}, the rearrangement term is explicitly given by
\begin{equation}
 \Sigma_R=-\frac{B_0}{3\Delta n}
 \frac{\exp\left(\dfrac{n-n'}{\Delta n}\right)}
 {\left[1+\exp\left(\dfrac{n-n'}{\Delta n}\right)\right]^2}.
\label{eq:WS_rearrangement}
\end{equation}

The kinetic contributions to the energy density, pressure, and baryon chemical potential in weak equilibrium are defined as:
\begin{align}
 \varepsilon_{\rm kin}&=\sum_q\frac{\gamma_q}{2\pi^2}\int_0^{k_{F,q}}
 k^2\sqrt{k^2+m_q^2}\,dk,
 \label{eq:energy_kin}\\
 p_{\rm kin}&=\sum_q\frac{\gamma_q}{6\pi^2}\int_0^{k_{F,q}}
 \frac{k^4}{\sqrt{k^2+m_q^2}}\,dk,
 \label{eq:pressure_kin}\\
 \mu_B^{\rm kin}&=\sum_q\sqrt{k_{F,q}^2+m_q^2}.
 \label{eq:muB_kin}
\end{align}
The thermodynamic quantities can then be written as:
\begin{align}
 \varepsilon&=\varepsilon_{\rm kin}+B(n),
 \label{eq:WS_energy}\\
 p&=p_{\rm kin}-B(n)+3n\Sigma_R,
 \label{eq:WS_pressure}\\
 \mu_B&=\mu_B^{\rm kin}+3\Sigma_R.
 \label{eq:WS_mu}
\end{align}
The equation of state for this particular parametrization is obtained simply by substituting Eqs.~\eqref{eq:woods_saxon} and \eqref{eq:WS_rearrangement} into Eqs.~\eqref{eq:WS_energy}--\eqref{eq:WS_mu}.

\begin{figure*}[ht]
\begin{tabular}{ccc}
\centering 
\includegraphics[scale=.58, angle=270]{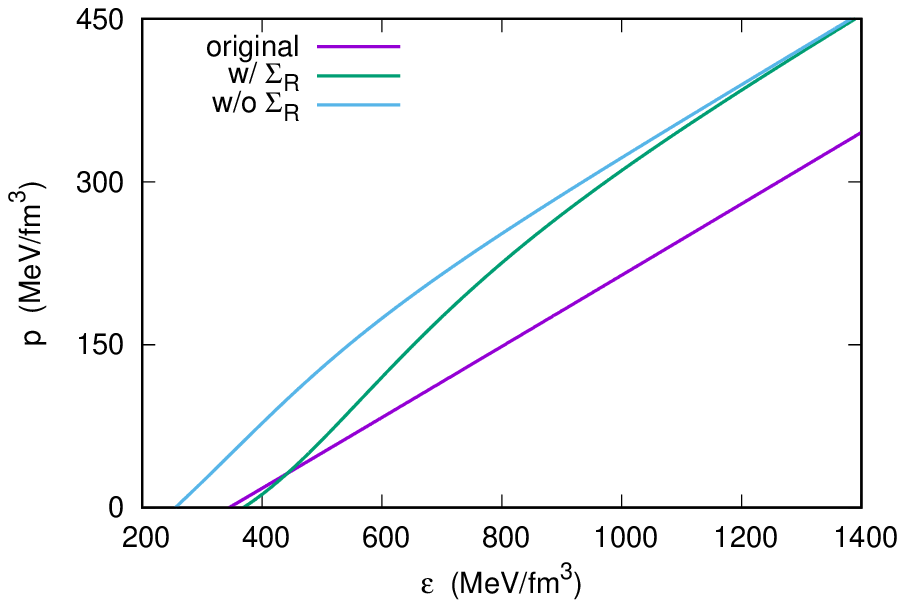} &
\includegraphics[scale=.58, angle=270]{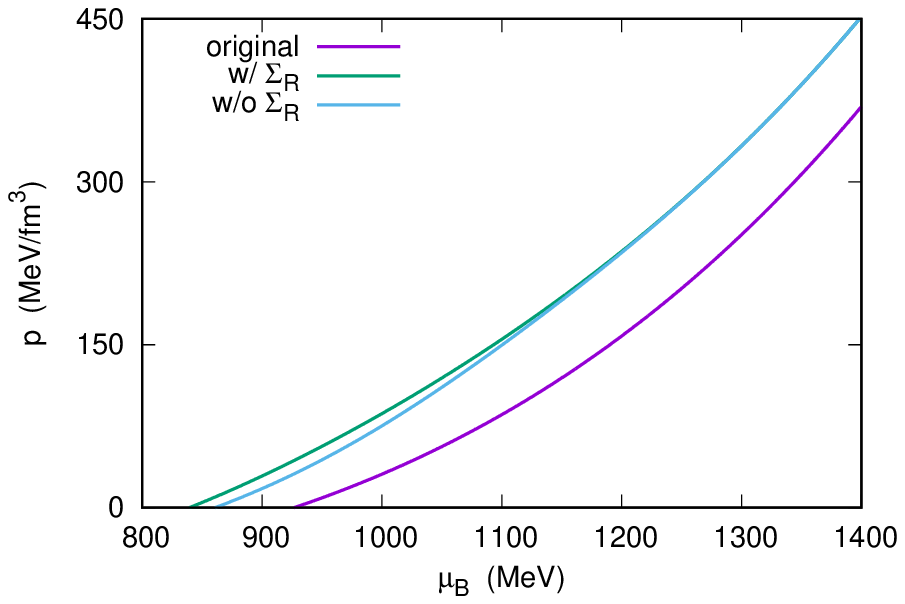} \\
\includegraphics[scale=.58, angle=270]{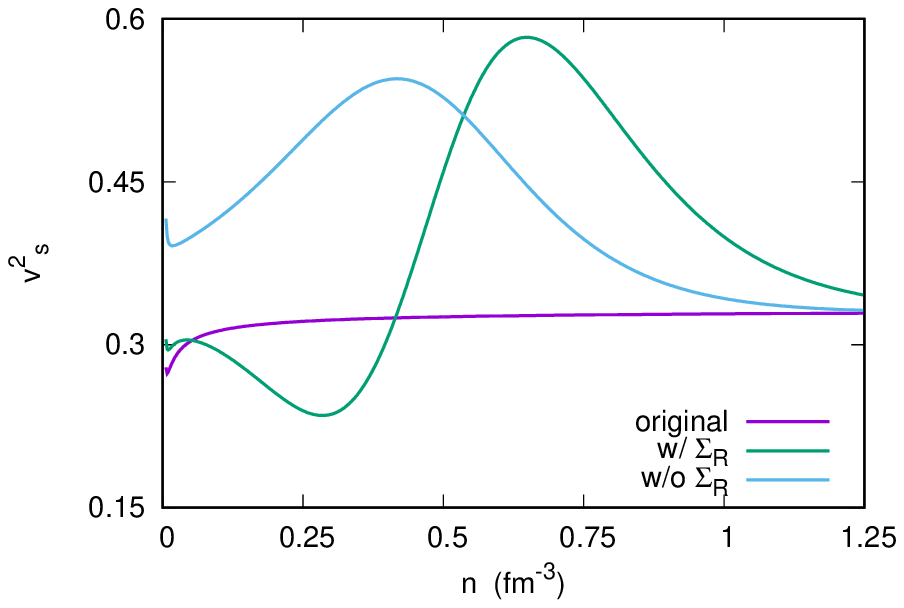} &
\includegraphics[scale=.58, angle=270]{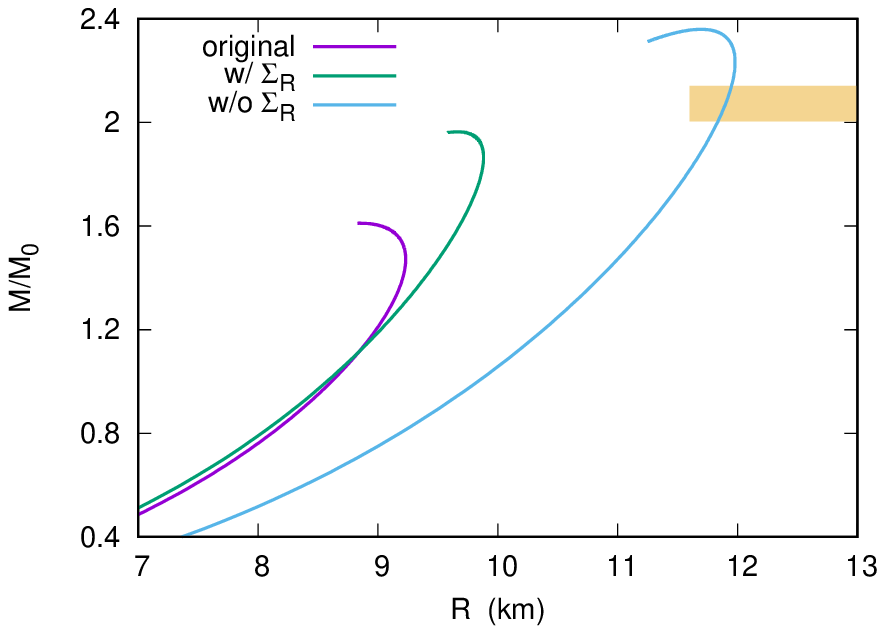}\\
\end{tabular}
\caption{The EOS (top-left), the pressure as a function of the chemical potential (top-right), the speed of sound (bottom-left), and TOV solutions (bottom-right) for the MIT bag model with a constant bag, a density-dependent bag model with and without rearrangement terms $\Sigma_R$. The hatched area is the inferred mass and radius of the  PSR J0740+6620 pulsar, as discussed in Ref.~\cite{Salmi_2024}.} \label{MIT}
\end{figure*}

We now present numerical results for the MIT bag model with a constant bag, and density-dependent bag models with and without rearrangement terms, $\Sigma_R$, for chemically stable matter, on which leptons are also added as a free Fermi gas.
In this particular case, I use $m_u=m_d=4$ MeV, and $m_s = 90$ MeV. 
I also use $n' = 0.45$ fm$^{-3}$, and $\Delta n = 0.15$ fm$^{-3}$, and $B_0^{1/4} = 159$ MeV.
This value of $B_0$ falls inside the stability window for both the constant and density-dependent bag model (see Refs.~\cite{lopesps1,lopesps2} and the references therein for additional details).  The results are presented in Fig.~\ref{MIT}.

Beginning with the equation of state (EOS), the constant-bag model produces the softest EOS, whereas the density-dependent (DD) model without rearrangement terms yields the stiffest one. The thermodynamically consistent DD model exhibits an intermediate behavior, effectively bridging these two descriptions. At high densities, however, it approaches the DD model without rearrangement terms, as $\Sigma_R$ gradually vanishes.

A second feature analyzed is the pressure as a function of the chemical potential. Although not discussed here, this is particularly important in the study of hybrid stars~\cite{Lopes2022ApJ}. We can see that the effect of the rearrangement terms is small in this case; yet, the thermodynamically consistent model predicts a large value of $p$ for a fixed $\mu$, indicating an earlier deconfinement when compared with the model without $\Sigma_R$.

A third physical quantity is the square of the speed of sound defined as:

\begin{equation}
 v^2_S = \frac{\partial p}{\partial \varepsilon}  . 
\end{equation}

The numerically calculated values of $v_s^2$ are presented in the bottom-left panel of Fig.~\ref{MIT}. For the constant-bag model, $v_s^2$ is approximately 0.3 at low densities and gradually approaches the conformal limit of $1/3$ as the density increases. In contrast, the thermodynamically consistent DD model exhibits a non-monotonic behavior, with a pronounced peak around $n=0.60$ fm$^{-3}$. Nevertheless, at sufficiently high densities, the speed of sound also approaches the conformal limit, $v_s^2\to 1/3$.
We can understand better the non-monotonic behavior of $v^2_s$ following Eq. (\ref{eq:WS_energy}) and Eq.~(\ref{eq:WS_pressure}) writing:

\begin{equation}
 v^2_s = \bigg (\frac{dp_{kin}}{dn} +3n\Sigma'_R \bigg )\bigg(\frac{d\varepsilon_{kin}}{dn} + 3\Sigma_R \bigg)^{-1},   
\end{equation}
and realize that $\Sigma_R'$ changes sign at $n =n'$. Finally, without the rearrangement terms, we have peculiar behaviour at low density: when $v^2_s$ is already above the conformal limit, it peaks and ultimately approaches the conformal limit as $\Sigma_R \to 0$ in the high-density limit.

The last figure presented in the bottom-right of Fig.~\ref{MIT} is the mass-radius solution obtained from the TOV solution~\cite{TOV}. Here, I also display as a hatched area the constraint related to the mass and radius of the PSR J0740+6620 pulsar. As pointed out in Ref.~\cite{Salmi_2024}, after analyzing 3.6 years of NICER data, the authors inferred a mass and radius of $2.073 \pm 0.069M_\odot$ and 12.49$^{+1.28}_{-0.88}$ km, respectively.  As shown, the thermodynamically consistent model discussed here does not satisfy this constraint, althougth it clearly stiffens the EOS.

\section{Chemical potential-dependent bag models}

At the chemical potentials relevant to compact stars,
uncertainties in the properties of dense quark matter motivate
a phenomenological description incorporating the leading
effects of strong interactions, finite quark masses, and
pairing. Following Ref.~\cite{Alford2005}, I adopt a
parametrization of the grand thermodynamic potential density
in terms of quartic and quadratic powers of the quark chemical
potential. The quartic coefficient accounts phenomenologically
for strong-interaction corrections to the free-quark Fermi sea,
while the quadratic term incorporates the leading mass and
color-superconducting contributions. For CFL matter, this
description retains corrections of order $m_q^2/\mu^2$ and
$\Delta^2/\mu^2$, neglecting higher-order terms
\cite{Lugones2002,Alford2005}. The accuracy of this truncation
therefore depends on these dimensionless ratios and improves
as they decrease. This parametrization provides a simple
framework for exploring quark matter beyond the noninteracting
MIT bag model and was also employed in
Ref.~\cite{Pereira2018} to investigate the dynamical stability
of hybrid stars. In the present work, I extend this framework
by allowing the bag contribution to depend explicitly on
the quark chemical potential.

In the conventional MIT bag model discussed in the last section, all three quark flavors are treated with finite masses. For consistency, I adopt the same prescription in the CFL model, retaining the physical quark masses rather than neglecting the up- and down-quark contributions. Accordingly, the leading quadratic mass correction includes all three flavors, instead of only the strange quark, as commonly assumed in the literature~\cite{Alford2005,Pereira2018}. Since $m_u$ and $m_d$ are much smaller than $m_s$, their numerical contributions are expected to be small, but their inclusion ensures a consistent treatment of quark masses throughout this work. At zero temperature, the grand thermodynamical potential can be defined as~\cite{Alford2005,Lugones2002,Pereira2018}:

\begin{equation}
 \Omega(\mu)=-\frac{3a_4}{4\pi^2}\mu^4
 +\frac{3a_2}{4\pi^2}\mu^2+B(\mu),
\label{eq:mu_omega}
\end{equation}
where $\mu$ is the common quark chemical potential and $B(\mu)$ is an arbitrary differentiable function. To leading order in the strong coupling and to quadratic order in the quark masses and the color--flavor-locked (CFL) pairing gap, the coefficients are~\cite{Buballa2005}:

\begin{align}
 a_4&=1-\frac{2\alpha_s}{\pi},
 \label{eq:mu_a4}\\
 a_2&=m_u^2+m_d^2+m_s^2-4\Delta^2.
 \label{eq:mu_a2}
\end{align}
Where $\alpha_s$ is the strong coupling constant and $\Delta$ is a single CFL pairing gap. Here, I extend the mass contribution by summing the leading correction $3m_q^2\mu^2/(4\pi^2)$ over the three flavors, instead of only $m_s$, as done in Refs.~\cite{Alford2005,Pereira2018}. The pairing contribution is $-3\Delta^2\mu^2/\pi^2$, and the equation~\eqref{eq:mu_omega} is a truncated effective description: mixed corrections proportional to $(\alpha_s m_q^2 \mu^2)$, higher-order perturbative terms, and terms of fourth and higher order in the masses or gap are not retained. In particular, $a_2$ has dimensions of energy squared and can be negative when pairing dominates. In the following derivatives, $\alpha_s$, the quark masses, and $\Delta$ are held fixed; only $B$ depends on $\mu$.

To preserve the notation of Sec.~\ref{sec:formalism}, $n$ continues to denote the baryon density. I denote the total quark number density by $n_Q= \sum_q n_q$, so that:

\begin{equation}
 n_Q=3n,\qquad \mu_B=3\mu,
\label{eq:mu_conventions}
\end{equation}
which guarantees:

\begin{equation}
 \mu n_Q = \mu_B n.
\end{equation}

The pressure follows directly from the grand thermodynamic potential density:
\begin{equation}
 p(\mu)=-\Omega(\mu)
 =\frac{3a_4}{4\pi^2}\mu^4
 -\frac{3a_2}{4\pi^2}\mu^2-B(\mu).
\label{eq:mu_pressure}
\end{equation}
The density conjugate to $\mu$ is the total quark density,
\begin{equation}
 n_Q(\mu)=-\frac{d\Omega}{d\mu}
 =\frac{3a_4}{\pi^2}\mu^3
 -\frac{3a_2}{2\pi^2}\mu-B'(\mu).
\label{eq:mu_quark_density}
\end{equation}

The contribution arising from the derivative of the bag function can be treated as the rearrangement contribution to the quark density: 

\begin{equation}
 \mathcal{N}_R\equiv-B'(\mu).
\label{eq:mu_rearrangement}
\end{equation}

Unlike $\Sigma_R$ in Eq.~\eqref{eq:rearrangement}, $\mathcal{N}_R$ has dimensions of number density. With this definition,
\begin{align}
 n_Q(\mu)&=\frac{3a_4}{\pi^2}\mu^3
 -\frac{3a_2}{2\pi^2}\mu+\mathcal{N}_R,
 \label{eq:mu_density_rearranged}\\
 n(\mu)&=\frac{a_4}{\pi^2}\mu^3
 -\frac{a_2}{2\pi^2}\mu+\frac{\mathcal{N}_R}{3}.
 \label{eq:mu_baryon_density}
\end{align}

Thus, a decreasing bag function gives a positive rearrangement contribution to the density.

The energy density can be derived independently of the Euler relation from the grand partition function. Writing $\beta=1/T$ and introducing the total quark number operator $\hat N_Q$, I define:

\begin{equation}
 \mathcal Z=\operatorname{Tr}
 \exp\left[-\beta\hat H+(\beta\mu)\hat N_Q\right],
 \qquad
 \Omega=-\frac{\ln\mathcal Z}{\beta V}.
\label{eq:mu_partition}
\end{equation}

For a Hamiltonian with no explicit dependence on the thermodynamic sources, differentiation at fixed $\beta\mu$ and $V$ gives
\begin{equation}
 \varepsilon=-\frac{1}{V}
 \left.\frac{\partial\ln\mathcal Z}{\partial\beta}\right|_{\beta\mu,V}
 =\left.\frac{\partial(\beta\Omega)}{\partial\beta}\right|_{\beta\mu,V}.
\label{eq:mu_energy_partition}
\end{equation}
The condition $\beta\mu=\mathrm{const.}$ implies $d\mu/d\beta=-\mu/\beta$. Applying the chain rule before taking the zero-temperature limit yields:

\begin{equation}
 \varepsilon=\Omega-T\frac{\partial\Omega}{\partial T}
 -\mu\frac{\partial\Omega}{\partial\mu}.
\label{eq:mu_energy_finite_temperature}
\end{equation}
For the zero-temperature potential in Eq.~\eqref{eq:mu_omega}, this becomes:

\begin{equation}
 \varepsilon(\mu)=\Omega(\mu)-\mu\frac{d\Omega}{d\mu}
 =-\mu^2\frac{d}{d\mu}\left(\frac{\Omega(\mu)}{\mu}\right),
\label{eq:mu_energy_derivative}
\end{equation}
where the last expression is understood for $\mu>0$. Explicit differentiation gives:
\begin{align}
 \varepsilon(\mu)&=\frac{9a_4}{4\pi^2}\mu^4
 -\frac{3a_2}{4\pi^2}\mu^2+B(\mu)-\mu B'(\mu)\notag\\
 &=\frac{9a_4}{4\pi^2}\mu^4
 -\frac{3a_2}{4\pi^2}\mu^2+B(\mu)+\mu \mathcal{N}_R.
\label{eq:mu_energy}
\end{align}
Consequently, the rearrangement contribution to the energy density is:

\begin{equation}
 \delta\varepsilon=\mu \mathcal{N}_R=-\mu B'(\mu).
\label{eq:mu_energy_rearrangement}
\end{equation}

In this formulation, $B(\mu)$ is an additive contribution to the grand thermodynamic potential density. Its contribution to the energy density is $B(\mu)-\mu B'(\mu)$. Specifying $B$ as a function of $\mu$ in Eq.~\eqref{eq:mu_omega} therefore defines a different starting ansatz from adding $B(n)$ to the energy density in Sec.~\ref{sec:formalism}.

I can now verify the Euler relation by adding the independently obtained expressions for the energy density and pressure. Equations~\eqref{eq:mu_pressure} and \eqref{eq:mu_energy} give:

\begin{align}
 \varepsilon+p
 &=\frac{3a_4}{\pi^2}\mu^4
 -\frac{3a_2}{2\pi^2}\mu^2+\mu \mathcal{N}_R\notag\\
 &=\mu\left[\frac{3a_4}{\pi^2}\mu^3
 -\frac{3a_2}{2\pi^2}\mu+\mathcal{N}_R\right]\notag\\
 &=\mu n_Q=\mu_B n.
\label{eq:mu_euler_check}
\end{align}

For nonzero density, the chemical potentials therefore satisfy:

\begin{equation}
 \mu=\frac{\varepsilon+p}{n_Q},
 \qquad
 \mu_B=\frac{\varepsilon+p}{n}.
\label{eq:mu_enthalpy_ratio}
\end{equation}
The differential consistency can also be checked. Wherever $\Omega$ is twice differentiable:

\begin{equation}
 \frac{dp}{d\mu}=n_Q,\qquad
 \frac{d\varepsilon}{d\mu}=-\mu\Omega''(\mu)
 =\mu\frac{dn_Q}{d\mu}.
\label{eq:mu_first_law_check}
\end{equation}
Hence $d\varepsilon=\mu\,dn_Q=\mu_B\,dn$. On a branch where the density can be used as an independent variable, Eqs.~\eqref{eq:mu_euler_check} and \eqref{eq:mu_first_law_check} also imply:

\begin{equation}
 n^2\frac{d(\varepsilon/n)}{dn}=n\mu_B-\varepsilon=p.
\label{eq:mu_pressure_consistency}
\end{equation}

These relations demonstrate the thermodynamic consistency of the model without imposing the Euler relation as the definition of the energy density. Mechanical stability and causality remain separate conditions to be checked for each choice of $B(\mu)$. For a constant bag function, $\mathcal{N}_R=0$ and the usual expressions of this truncated model are recovered.

As an explicit example, I again choose a Woods-Saxon-type Bag model:

\begin{equation}
 B(\mu)=\frac{B_0}{1+\exp\left(\dfrac{\mu-\mu_0}{w}\right)},
\label{eq:mu_woods_saxon}
\end{equation}
where $B_0>0$ has dimensions of energy density, $\mu_0$ is the quark chemical potential at which $B=B_0/2$, and $w>0$ controls the width of the transition. Differentiating Eq.~\eqref{eq:mu_woods_saxon} gives:
\begin{equation}
 B'(\mu)=-\frac{B_0}{w}
 \frac{\exp\left(\dfrac{\mu-\mu_0}{w}\right)}
 {\left[1+\exp\left(\dfrac{\mu-\mu_0}{w}\right)\right]^2}.
\label{eq:mu_WS_derivative}
\end{equation}
The rearrangement contribution is therefore:

\begin{align}
 \mathcal{N}_R&=\frac{B_0}{w}
 \frac{\exp\left(\dfrac{\mu-\mu_0}{w}\right)}
 {\left[1+\exp\left(\dfrac{\mu-\mu_0}{w}\right)\right]^2}
\label{eq:mu_WS_rearrangement} .
\end{align}

The corresponding energy contribution is $\mu \mathcal{N}_R$. Substituting Eqs.~\eqref{eq:mu_woods_saxon} and \eqref{eq:mu_WS_rearrangement} into Eqs.~\eqref{eq:mu_pressure}, \eqref{eq:mu_baryon_density}, and \eqref{eq:mu_energy} gives the equation of state parametrically in $\mu$, including all contributions from the chemical-potential dependence of the bag function.

\begin{figure*}[ht]
\begin{tabular}{ccc}
\centering 
\includegraphics[scale=.58, angle=270]{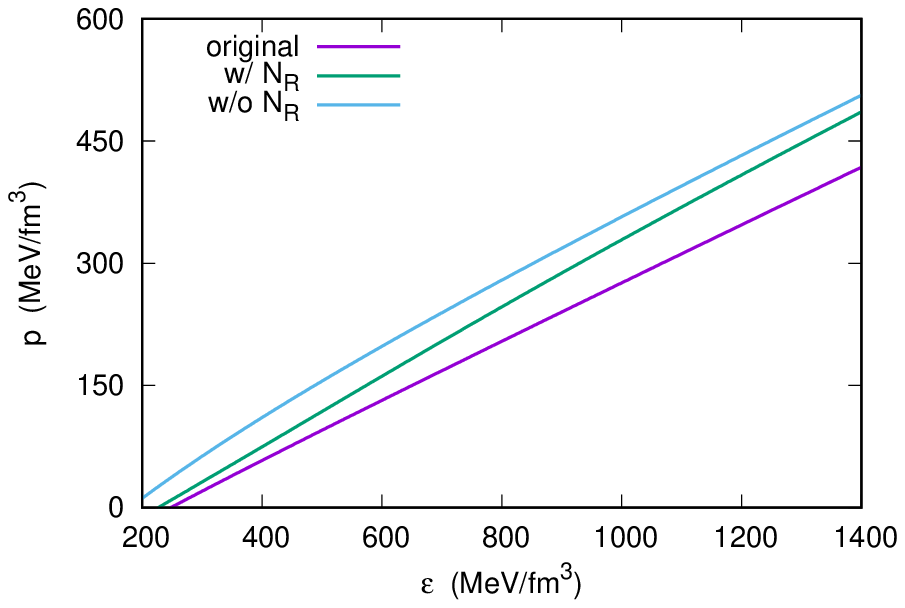} &
\includegraphics[scale=.58, angle=270]{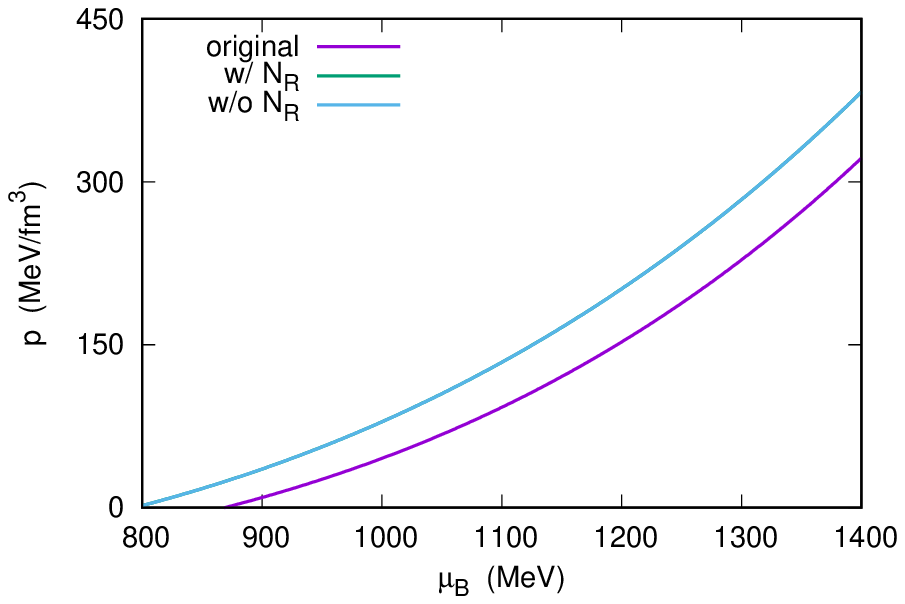} \\
\includegraphics[scale=.58, angle=270]{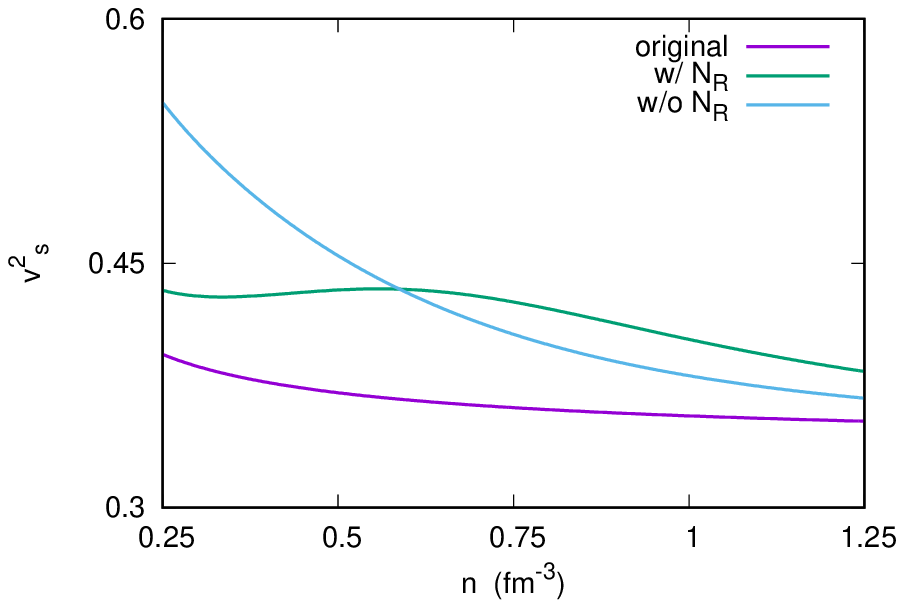} &
\includegraphics[scale=.58, angle=270]{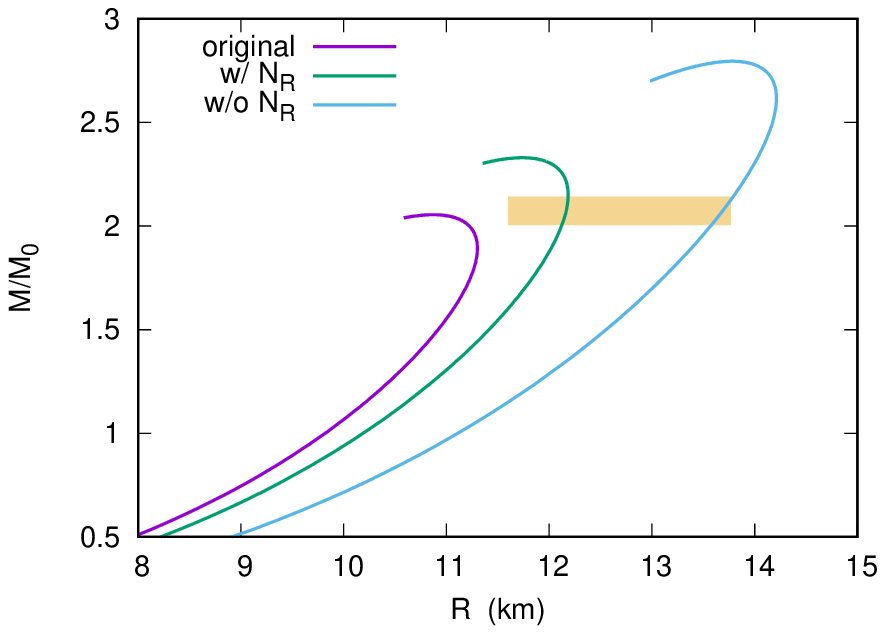}\\
\end{tabular}
\caption{The EOS (top-left), the pressure as a function of the chemical potential (top-right), the speed of sound (bottom-left) and TOV solutions (bottom-right) for the perturbative CFL bag model with a constant bag, a chemical-potential-dependent bag model with and without rearrangement terms $\mathcal{N}_R$. The hatched area is the inferred mass and radius of the  PSR J0740+6620 pulsar, as discussed in Ref.~\cite{Salmi_2024}.} \label{CFL}
\end{figure*}

For a numerical analyzes, in Fig.~\ref{CFL} I display the properties with $a_4 = 0.7$ and $\Delta = 100$ MeV (as suggested in Ref.~\cite{Lugones2002}), $\mu_0 = 350$ MeV, and $w = 80$ MeV. I also use $B_0^{1/4} =155$ MeV. This value implies that the strange quark matter is inside the stability window for the constant and the thermodynamically consistent bag model.  In the CFL phase, leptons are not present.

Starting with the EOSs, we observe a behavior similar to that found in the density-dependent case. The constant-bag model produces the softest EOS, whereas the model without rearrangement terms yields the stiffest one. Once again, the thermodynamically consistent model acts as a bridge between these two descriptions, although the curves with and without rearrangement terms approach each other only at much higher densities.

The next feature is the pressure as a function of the chemical potential. Here, the constant bag presents the lowest pressure for a fixed chemical potential. However, the pressure for a non-constant bag is independent of the rearrangement terms. This happens because in a chemical-potential-dependent bag model, neither the pressure nor the chemical potential has rearrangement terms. Curiously, changing the dependence of the bag function from density to chemical potential reverses which thermodynamic quantities acquire rearrangement contributions. For a density-dependent bag function, the quark number densities retain their usual expressions in terms of the Fermi momenta, and the energy density retains the form $\varepsilon=\varepsilon_{\rm kin}+B(n)$. The quark chemical potentials and pressure, however, acquire the rearrangement contributions related to $\Sigma_R$. On the other hand, for a chemical-potential-dependent bag function, the chemical potential remains the independent thermodynamic variable, and the pressure retains its form, while the rearrangement contributions ($\mathcal{N}_R$) appear in the total quark number density and energy density.

In the bottom-left panel, I display the square of the speed of sound. Although not shown, $v_s^2$ increases sharply at very low densities. For the constant-bag model, $v_s^2\to1$ as $n\to0$, since the quadratic term associated with the pairing gap $\Delta$ dominates the derivatives of the kinetic contributions. For the chemical-potential-dependent bag models, the behavior is more singular, and superluminal values can arise at sufficiently low densities. Nevertheless, these results occur outside the domain of validity of Eq.~(\ref{eq:mu_omega}), which requires sufficiently small values of $(\Delta/\mu)^2$. Furthermore, these low-density regions correspond to negative pressures and are not reached in the stellar configurations considered here. For strange stars, $(\Delta/\mu)^2\approx0.14$ at the surface ($p=0$), decreasing to approximately $0.06$ at the stellar center.

For the constant-bag model and the model without rearrangement terms, $v_s^2$ decreases monotonically with density in the stellar regime, whereas the thermodynamically consistent model exhibits a peak around $n=0.70$ fm$^{-3}$. The thermodynamically consistent speed of sound can be written as:

\begin{equation}
v_s^2 =
\bigg(
\frac{3a_4}{\pi^2}\mu^3
-\frac{3a_2}{2\pi^2}\mu + N_R
\bigg)
\bigg(
\frac{9a_4}{\pi^2}\mu^3
-\frac{3a_2}{2\pi^2}\mu + \mu N_R'
\bigg)^{-1}. \\ 
\end{equation}

It is clear that for the Woods-Saxon type bag model, $v^2_S\to1/3$ at the high-density limit, although much more slowly than the traditional MIT bag model due to the factor $a_2$.

Finally, in the bottom-right, I display the TOV solution together with the constraint coming from the PSR J0740+6620 pulsar. The paring gap, $\Delta = 100$ MeV, is already enough to sustain a 2.05 $M_\odot$ strange star. However, it fails to satisfy the constraint coming from the  PSR J0740+6620 pulsar because its radius is below 11.6 km. On the other hand, the thermodynamically consistent model in the CFL phase can reach a maximum mass of 2.33$M_\odot$ and radii consistent with the observational constraints for PSR J0740+6620. Therefore, it was possible to simulate massive strange stars whose speed of sound satisfies the conformal limit $v^2_s\to 1/3$ at the high-density limit. Lastly, a maximum mass of 2.80$M_\odot$ is reached without the rearrangement terms. This suggests that neglecting the rearrangement terms may lead to a significant overestimation of the maximum stellar mass.

\begin{figure}[ht]
  \begin{centering}
\begin{tabular}{c}
\includegraphics[width=0.334\textwidth,angle=270]{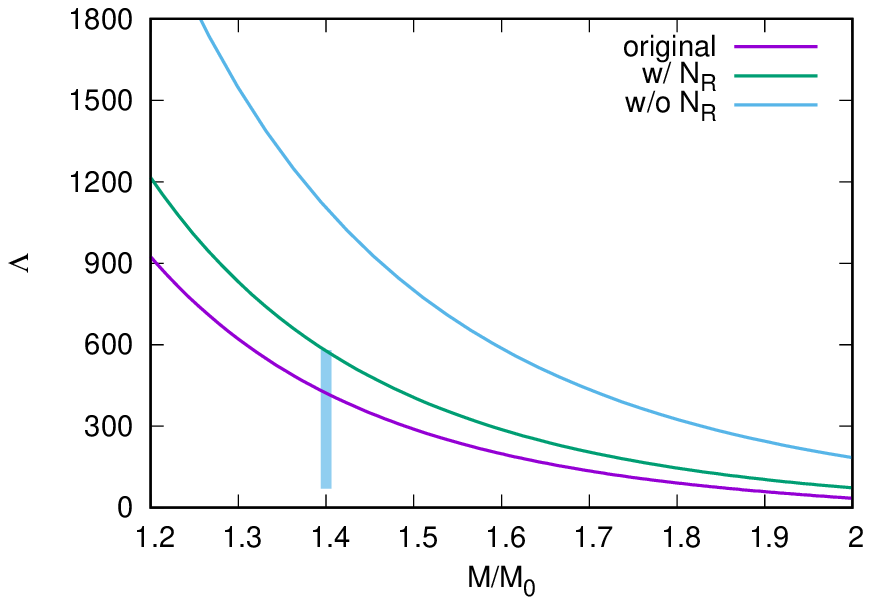} \\
\end{tabular}
\caption{Dimensionless tidal parameter $\Lambda$ for three different approaches to the CFL phase EOS. The hatched area is the constraint related to the GW170817 event~\cite{AbbottPRL}.} \label{tid}
\end{centering}
\end{figure}

A final test is the evaluation of the dimensionless tidal parameter $\Lambda$. If we put an extended body in an inhomogeneous external field, it will experience different forces throughout its extent. The result is a tidal interaction. The tidal deformability of a compact object quantifies how easily the object is deformed when subjected to an external tidal field. A larger tidal deformability indicates that the object is easily deformable. On the opposite side, a compact object with a smaller tidal deformability parameter is smaller, more compact, and it is more difficult to deform it. The dimensionless tidal deformability parameter $\Lambda$ is defined as:

\begin{equation}
 \Lambda = \frac{2k_2}{3C^5}   ,
\end{equation}
where M is the compact object’s mass and $C = GM/R$ is its compactness. The parameter $k_2$  is called Love number and is related to the metric perturbation. A complete discussion about the Love number and its calculation is both, very extensive and well documented in the literature and the interested reader can check Refs.~\cite{Hinderer_2008,tidal2010,Chat2020,Flores2020,AbbottPRL} and the references therein. In the case of the strange stars, the finite value of the energy density at the stars' surface must be taken into account, as discussed in Refs.~\cite{Odilon2021,Lopes2022ApJ}.

Gravitational-wave observations, such as those made by LIGO/Virgo, provide an additional probe of the dense-matter EOS through the tidal deformability of compact stars. In particular, the analysis of GW170817 constrains the dimensionless tidal deformability of a $1.4\,M_\odot$ star to $70<\Lambda_{1.4}<580$~\cite{AbbottPRL}. I now test the three EOSs related to the CFL phase in light of this constraint. The results are displayed in Fig.~\ref{tid}.

The thermodynamically consistent model, with the rearrangement terms, presents $\Lambda_{1.4} = 578$, which is very close but still lower than the upper limit of 580 inferred in Ref.~\cite{AbbottPRL}. On the other hand, the very massive model without the rearrangement terms has $\Lambda_{1.4} = 1098$, far above the upper limit. The CFL model with a constant bag is also in agreement with the GW170817 event, but fails to fulfill the constraint related to the  PSR J0740+6620 pulsar.

\section{Conclusions}

Motivated by recent applications of non-constant bag models in which thermodynamic consistency has not been explicitly verified~\cite{Sen2021,Pal2023,Podder2024,Sedaghat2025,Bordbar2012}, I have derived consistent formulations for quark matter with density-dependent and chemical-potential-dependent bag functions. In both cases, I have obtained the corresponding rearrangement contributions and explicitly demonstrated that the resulting thermodynamic quantities satisfy both the first law of thermodynamics and the Euler relation.

QCD predicts asymptotic freedom at sufficiently high densities, motivating a description that accounts for the evolution of nonperturbative vacuum effects in dense matter. A constant bag parameter fails to describe the running coupling constant of QCD, although its relative importance becomes negligible in the asymptotic limit. Allowing this parameter to decrease with density or chemical potential provides a phenomenological way to describe medium effects. Suitable bag functions can stiffen the EOS at stellar densities while recovering the conformal limit of the sound speed, $v_s^2\rightarrow1/3$, as their derivative contributions become asymptotically negligible. This behavior differs from that of conventional models with a constant repulsive vector coupling and no vector self-interactions, in which the vector contribution eventually dominates and drives $v_s^2$ toward unity.

I first considered a bag function $B(n)$ depending on the baryon number density. The quark number densities retain their usual expressions in terms of the Fermi momenta, and the energy density is $\varepsilon=\varepsilon_{\mathrm{kin}}+B(n)$, with no separate rearrangement contribution. Differentiating this energy density with respect to the individual quark number densities yields the rearrangement term $\Sigma_R=\frac{1}{3}dB/dn$, which has dimensions of energy. This term shifts each quark chemical potential by $\Sigma_R$ and contributes $3n\Sigma_R$ to the pressure. For the parametrization studied, the thermodynamically consistent model stiffened the EOS while preserving the asymptotic conformal limit, $v_s^2\rightarrow1/3$. The squared sound speed exhibited a non-monotonic behavior, with a peak near $n=0.60$ fm$^{-3}$. Nevertheless, the corresponding stellar sequence did not satisfy the adopted mass–radius constraint for PSR J0740+6620.

I then extended the study to color-superconducting quark matter in the CFL phase, described by a grand thermodynamic potential expressed in terms of the quark chemical potential. Allowing the bag parameter to depend on $\mu$, I constructed a thermodynamically consistent formulation in which the rearrangement term $\mathcal{N}_R=-dB/d\mu$ has dimensions of number density. In this case, the pressure retains the form $p=-\Omega$, and the chemical potential remains the independent thermodynamic variable. The total quark number density and the energy density acquire the rearrangement contributions $\mathcal{N}_R$ and $\mu\mathcal{N}_R$, respectively. Color superconductivity provides an additional source of stiffness. For the parametrization studied, the resulting thermodynamically consistent model satisfies the adopted mass–radius constraint for PSR J0740+6620~\cite{Salmi_2024} while recovering the conformal limit, $v_s^2\rightarrow1/3$, at asymptotically high densities. The approach to this limit is slower than in the unpaired MIT bag model, since the pairing gap keeps the squared sound speed above $1/3$ over a broader density range.

Finally, I tested the thermodynamically consistent CFL model against the tidal-deformability constraint inferred from GW170817, detected by the LIGO/Virgo gravitational-wave detectors. The resulting value, $\Lambda_{1.4} =578$, lies within the adopted interval, $70<\Lambda_{1.4}<580$~\cite{AbbottPRL}. This model is therefore particularly promising, as it simultaneously satisfies the adopted astrophysical constraints on stellar masses, radii, and tidal deformability while recovering the conformal limit of the sound speed at asymptotically high densities~\cite{Kurkela2010}.

This study represents a first step toward exploring the implications of thermodynamically consistent non-constant bag models. The analysis can be extended to hybrid stars, finite-temperature effects, magnetized quark matter, and other parametrizations of the bag functions. Further work along these directions is currently in progress.

\vspace{0.75cm}

\textbf{Acknowledgments:} L.L.L.  was partially supported by CNPq (Brazil)
under Grant No 305347/2024-1, and INCT-FNA Project 408419/2024-5.

\bibliography{aref}

\end{document}